\documentclass[fleqn,usenatbib]{mnras} 

\usepackage[T1]{fontenc}
\usepackage{ae,aecompl}
\usepackage{mathptmx}

\usepackage{graphicx}	
\usepackage{amsmath}	
\usepackage{amssymb}	
\usepackage{subfigure}
\usepackage{hyperref}

\title[Orbital Evolution of Neptune's ring arcs]
{Neptune's ring arcs confined by coorbital satellites: dust orbital evolution through solar radiation}

\author[Giuliatti Winter et. al]{
S. M. Giuliatti Winter\thanks{E-mail: giuliatti.winter@unesp.br}, 
G. Madeira\thanks{E-mail: gustavo.o.madeira@unesp.br}  
and R. Sfair\thanks{E-mail: rafael.sfair@unesp.br}
\\
S\~ao Paulo State University -UNESP,  
Grupo de Din\^amica Orbital e Planetologia,
Guaratinguet\' a, CEP 12516-410, Brazil
}

\date{Accepted XXX. Received YYY; in original form ZZZ}

\pubyear{2015}

\begin{document}
\label{firstpage}
\pagerange{\pageref{firstpage}--\pageref{lastpage}}
\maketitle


\begin{abstract}

Voyager~2 images   confirmed  the presence of  ring arcs  around Neptune. 
 These structures 
need a confinement mechanism to constrain their spreading  due to collisions, dissipative forces, and differential
keplerian 
motion. Here
 we report the results of a set of numerical simulations of the system formed by Neptune, 
the satellite Galatea, dust ring particles and hypothetical co-orbital satellites. 
This dynamical system depicts a recent confinement 
mechanism formed  by  four co-orbital satellites being  responsible for the azimuthal  confinement of the arcs, while  
Galatea responds for  their radial confinement.  After the numerical simulations, the particles were divided  into four groups:  particles that stay in the arcs, transient particles, particles that leave the arcs to the Adams ring and particles that collide with the co-orbital satellites. Our results showed that 
in  all  arcs the lifetime of the  smaller particles  is at most 50~years. After 100~years  about $20\%$ of the total amount  of larger particles is still present in the arcs. 
 From our numerical simulations, the  particles should be  present  in all   arcs after 30~years,  the period  between  the discovery of the arcs up to now.  Our results can not explain the disappearance of the leading arcs,  Libert\'e and Courage, unless the arcs are  formed by different  particle sizes.  Analysis of the dust production, due to collisions between interplanetary debris onto the surface of the 
co-orbital satellites, ruled out the hypothesis that small satellites close to or in the arc 
structure could be its source.
\end{abstract}

\begin{keywords}
planets and satellites: rings,  dynamical evolution and stability
\end{keywords}



\section{Introduction}

The arcs, previously detected by stellar occultation in 1984 \citep{Si91}, are the
densest parts of the tenuous Adams ring \citep{Hu86}. The  Libert\'e,
Egalit\'e and Fraternit\'e  arcs have azimuthal widths of about 4$^{\circ}$, 4$^{\circ}$ and 10$^{\circ}$, 
respectively, while the  Courage  arc has
2$^{\circ}$ of azimuthal width; all of them have about 15km of radial width.

Corotation and Lindblad resonances are essential to the confinement mechanism.  The corotation
resonance prevents the arc of an azimuthal spreading, while the   Lindblad resonance prevents
its radial spreading.  A  previous analysis considered the satellite Galatea as the shepherd satellite of
the four arcs \citep{Po91}.  The  84:86 corotation inclined resonance   and the  42:43 outer Lindblad resonance
kept the radial and azimuthal widths of the arcs. The corotation inclined resonance (CIR)   induces the
formation of 84 equilibrium points, although only five of them are populated by   arc particles,
the  Fraternit\'e  arc occupies two sites.  Further data on the mean motion of the arcs showed
that their  semi-major axis  is  displaced from the corotation resonant semi-major axis \citep{Na02},
leaving the arcs without   azimuthal confinement.

\cite{Fo96}   analysed the coupling between three resonances with Galatea  ( 84:86 CIR,  84:86   Lindblad inclined  resonance (LIR),  and 42:43  Lindblad eccentric resonance (LER)) and its effect on the particles motion. They obtained  that the eccentricity forced by the   LER results in collisions with relative velocities on the order of 1m/s that drive out larger arc  particles. These particles are believed to be the source of the dust observed in the arcs. They also noticed that   solar radiation pressure removes dust particles from the corotation sites in  a few decades.

 Recent observations obtained from the Keck~II and the Hubble Space  Telescope
showed that the arcs have changed both in location and brightness since  their discovery. All the
arcs may have decayed in intensity, and the   Libert\'e and Courage arcs have almost disappeared \citep{Pa05,Sh13}.

A recent confinement model proposes that the Adams ring has a collection of small co-orbital
satellites located in specific positions \citep{Re14}. These co-orbital satellites  would be responsible
for maintaining the  arc particles in  stable azimuthal positions, while Galatea prevents their radial
spreading.  In detail, this model consists of four small co-orbital satellites ($S_1$, $S_2$, $S_3$ and $S_4$),
where $S_2$, $S_3$ and $S_4$ are azimuthally placed at longitudes $\Theta = 48^{\circ}$, 
$59^{\circ}$ and $72^{\circ}$, respectively,
from  $S_1$ (at $\Theta = 0^{\circ}$). The equilibrium 
positions of   $S_2$, $S_3$ and $S_4$ are close to the Lagrangian point $L_4$ (or $L_5$) of the 
satellite $S_1$. Besides their specific  positions,  to guarantee the nominal
location of the arcs in the Adams ring, their masses were accurately calculated to keep them in
stable positions \citep{Re14}. For a density of 1 g/cm$^3$, the larger co-orbital satellite has
a radius of 5.2km and the smallest one has a radius of 1.1km \citep{Re14},  all  below the
detection  limit  of the Voyager instruments and the recent  telescopes' data.

 A recent paper by \cite{He19} studied the dynamics of multiple massive bodies in  corotation resonance through numerical simulations. Their results showed that the bodies exchanged angular momentum and energy during the encounter which changes their orbits. Although it does not mean that if one body moves closer to the resonance the other  would move further from the resonance. They explained that this occurs because the timescale of the close encounters is short when compared with the synodic period of the particles and the body. They argued that the exchange in the energy  may be similar to a collisional system, although a detailed  investigation is necessary.
  
 In the current  work is   studied  the orbital evolution of a set of arc dust particles  to estimate 
their  lifetime. 
Section~\ref{grav} deals with the gravitational effects on the  arc particles, no dissipative force is included in the system.  In section~\ref{dissipative} we discuss the orbital  behaviour  of the particles under the solar radiation force, while in section~\ref{production} we 
analyse the dust production rate and its role as a source of  the  arc particles.  In the last section, we discuss our results.

\section{Orbital evolution of the arc particles}\label{grav}

In this section, we analyse the evolution of the  arc particles  assuming that  Galatea is  in a circular and equatorial orbit (case~1) and  in an  inclined eccentric orbit (case~2).

The dynamical  model evokes the presence of four small co-orbital 
satellites ($S_1$, $S_2$, $S_3$ and $S_4$),
where $S_2$, $S_3$ and $S_4$ are azimuthally located at longitudes $\Theta = 48.31^{\circ}$, 
$59.38^{\circ}$ and $72.19^{\circ}$, respectively,
from  $S_1$ (at $\Theta = 0^{\circ}$) (see Table~\ref{initial1}). 
 Their masses are given in Table~\ref{initial1}.

The  system is  formed by  Neptune, the satellite Galatea, four hypothetical co-orbital satellites and a sample of  ring arc particles. 
The parameters of  Neptune (mass $m$, radius $r$ and the gravitational 
coefficients $J_2$ and $J_4$) were taken from \cite{Ow91}. 
The orbital elements and  mass of Galatea are given in Table~2. It is initially in an eccentric ($e=0.00022)$ and  inclined orbit ($I=0.0231^\circ $) at semi-major axis 61953km. 

The co-orbital satellites  and the arc system are initially  at the same semi-major axis (62932.7km). The  eccentricity and inclination of the arcs  are  $e = 3 \times 10^{-4}$ and $I=0$ (Table~\ref{initial2}). 
 
 The longitude of  the  pericentre ($\varpi$) of the co-orbitals satellites and the arc particles   is $50.82^\circ$, the same value as adopted  by \cite{Po91}. The true longitude of  the  Fraternit\'e arc is $251.88^{\circ}$.
 
 Each arc is formed by a sample of 2000 particles randomly chosen about 1km from the semi-major axis of the  co-orbital satellites and $1^\circ$ of the angular positions $40^\circ$, $50^\circ$, $66.5^\circ$ and $84^\circ$. 
 
\begin{table*}
\centering
\caption{Longitude $\Theta$ and mass of the co-orbital satellites and  the arc particles.}
\label{initial1}
\begin{tabular}{lllllllll}
\hline\noalign{\smallskip}
 & $S_1$ & $S_2$ & $S_3$ & $S_4$ & Fraternit\'e & Egalit\'e & Libert\'e & Courage\\
\hline\noalign{\smallskip}
 $\Theta$ ($^{\circ}$) & 0.00 & 48.31 & 59.38 & 72.19 & 40.00 & 50.00 & 66.50 & 84.00 \\
mass ($10^{13}$ kg) &  60.00 & 0.54 & 1.17 & 0.66 & -- & -- & -- & -- \\
\hline\noalign{\smallskip}
\end{tabular}
\end{table*}

\begin{table*}

\centering
\caption{Orbital elements and mass of Galatea and the Fraternit\'e arc \citep{Ow91,Po91,Re14,showalter2019seventh}.}
\label{initial2}
\begin{tabular}{llllrrrl}
\hline\noalign{\smallskip}
& a (km) & e ($10^{-4}$) & I ($^{\circ}$) & $\varpi$ ($^{\circ}$) & $\Omega$ ($^{\circ}$) & $\lambda$ ($^{\circ}$) &  m ($10^{18}$ kg)\\
\hline\noalign{\smallskip}
Galatea & 61953.0 & 2.2 & 0.0231 & 225.81 & 196.94 &  351.114 & 1.94 \\
Fraternit\' e arc &  62932.7 & 3.0 & 0.0 & 50.82 & 0.0 & 251.88 & -- \\
\hline\noalign{\smallskip}
\end{tabular}
\end{table*}

Numerical simulations were 
performed using the variable timestep Burlish-St\"oer algorithm from Mercury package \citep{Ch99}   for a timespan of 1000~years. 
The algorithm presented in \cite{Re06} was used to transform the state vector into geometric orbital 
elements, in order  to avoid the short-period terms caused by  $J_2$ and $J_4$.

\begin{figure*}
\subfigure[]{\quad \includegraphics[height=0.15\paperheight]{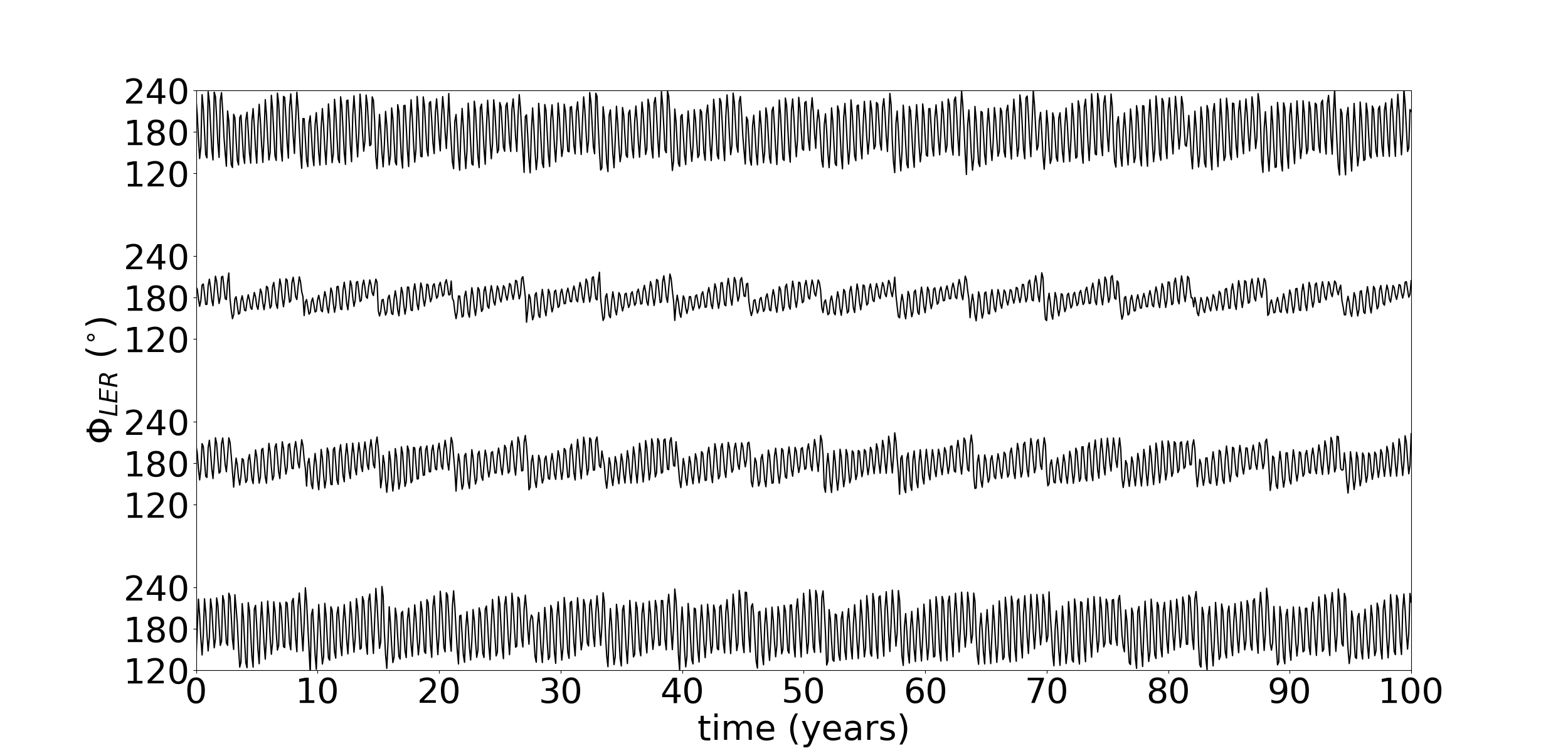}}
\subfigure[]{\quad \includegraphics[height=0.15\paperheight]{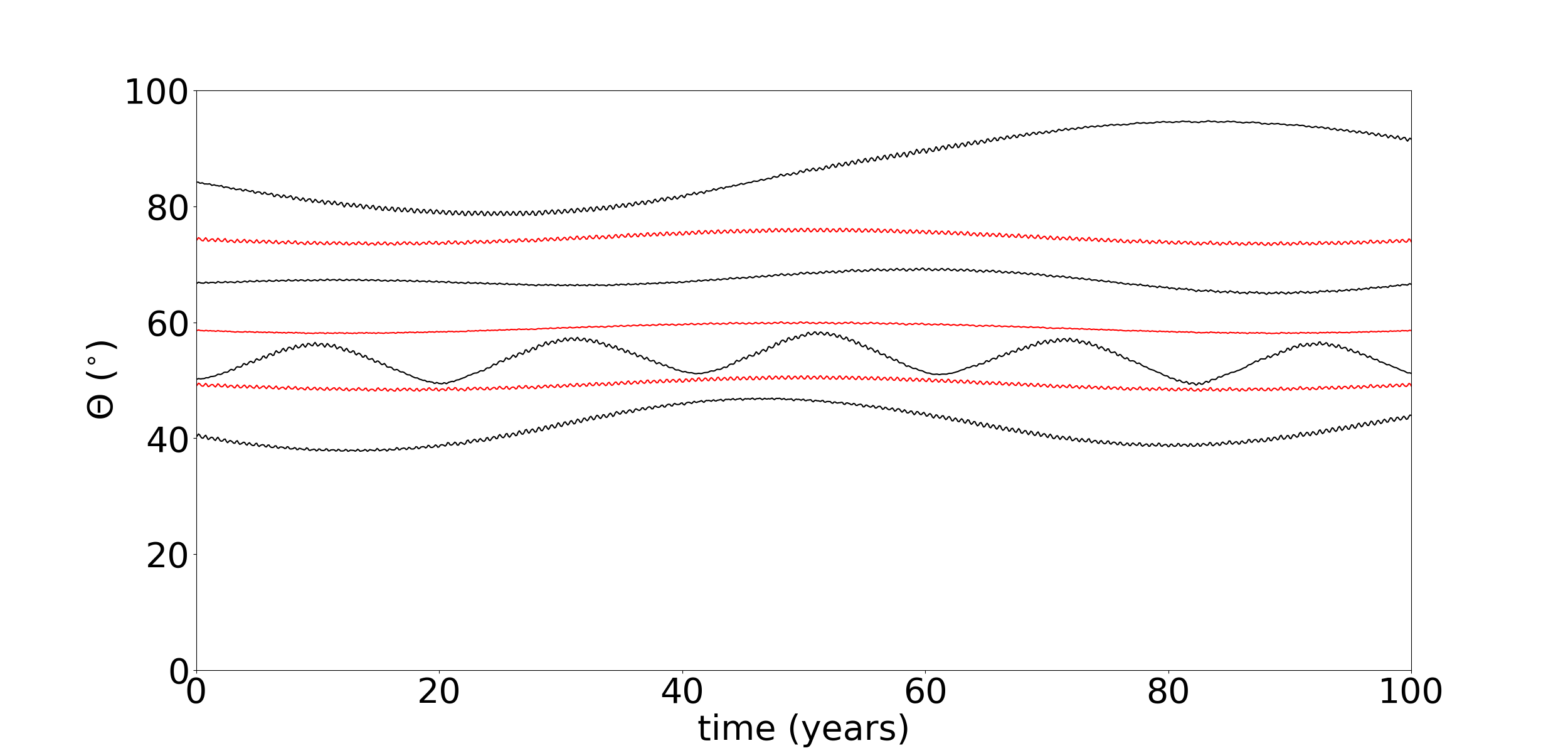}}
\caption{(a) $\Phi_{LER}$ and (b) $\Theta$ versus time for  representative particles at $\Theta=40^{\circ}$, $\Theta=50^{\circ}$, $\Theta=66.5^{\circ}$ and $\Theta=84^{\circ}$ from bottom to top. The particles are trapped in the 42:43 Lindblad eccentric resonance with Galatea and they are also azimuthally confined between the co-orbitals.}
\label{parts}
\end{figure*}

With these adopted initial conditions the co-orbital satellites 
and the  arc particles are trapped  in the 42:43 Lindblad resonance with Galatea.  The  resonant
angle is 
$\Phi_{\rm LER}=43\lambda - 42\lambda_{\rm G} - \varpi$, where $\lambda$ and $\varpi$ are  the mean longitude 
and the 
 longitude of the pericentre of the co-orbital satellites/particles, respectively, and $\lambda_{\rm G}$ is the mean 
longitude of Galatea.  Figure~\ref{parts} shows the resonant  and azimuthal angles (longitude of the particle in the frame rotating with the mean motion of the co-orbital $S_1$, $\Theta=\lambda-\lambda_{S_1}$) librating for representative  particles of each  arc.

\begin{figure}
\subfigure[]{\quad \includegraphics[height=0.135\paperheight]{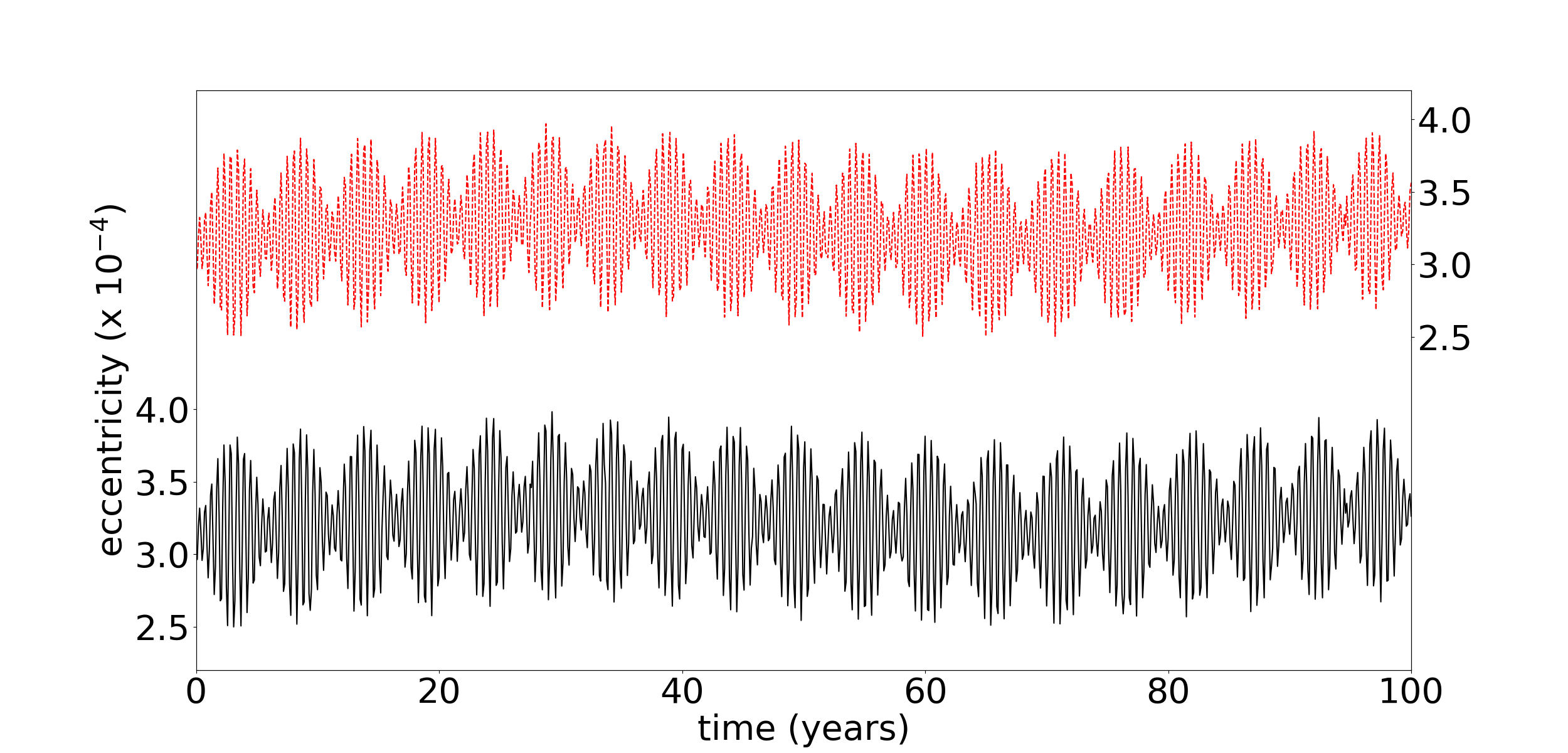}}
\subfigure[]{\quad \includegraphics[height=0.135\paperheight]{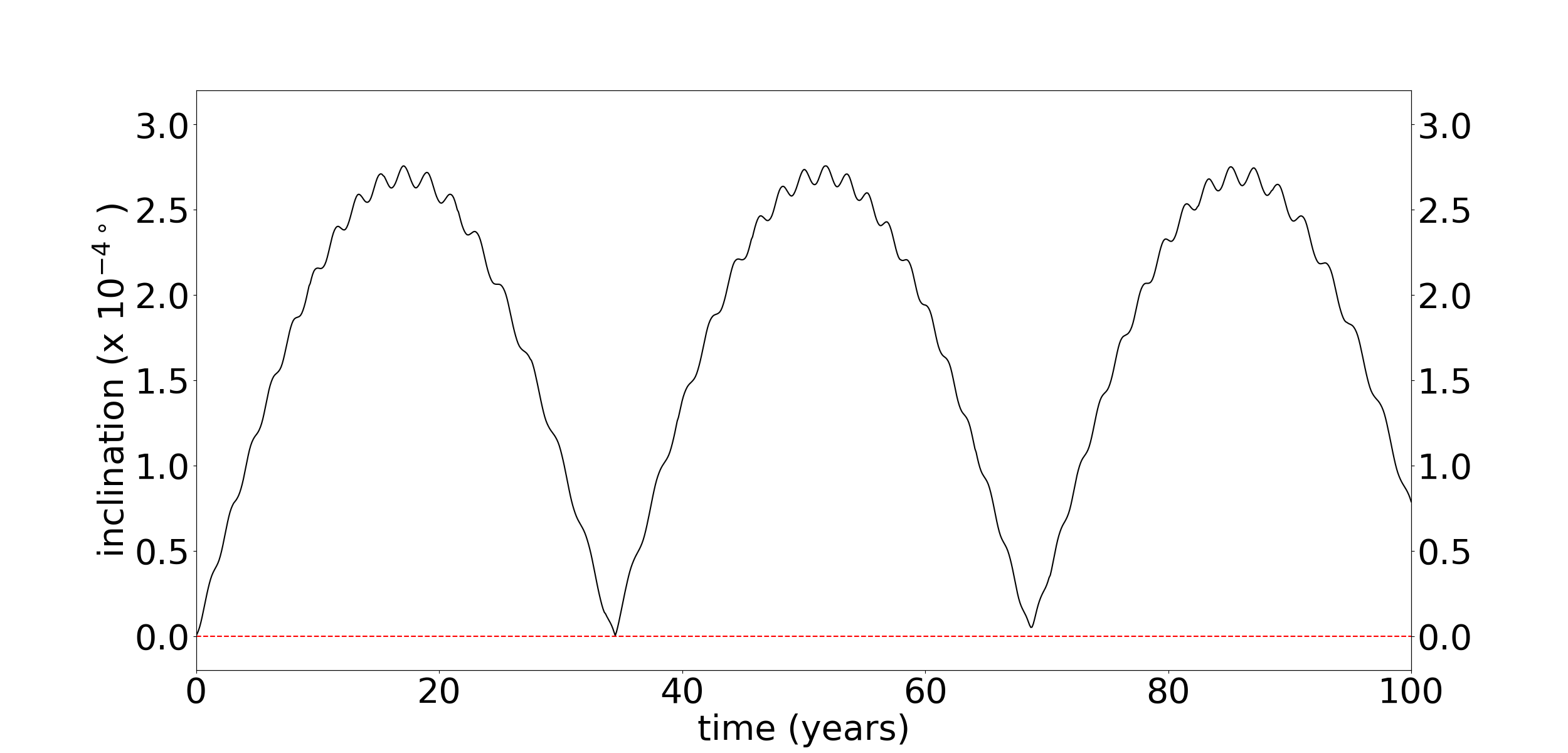}}
\subfigure[]{\quad \includegraphics[height=0.135\paperheight]{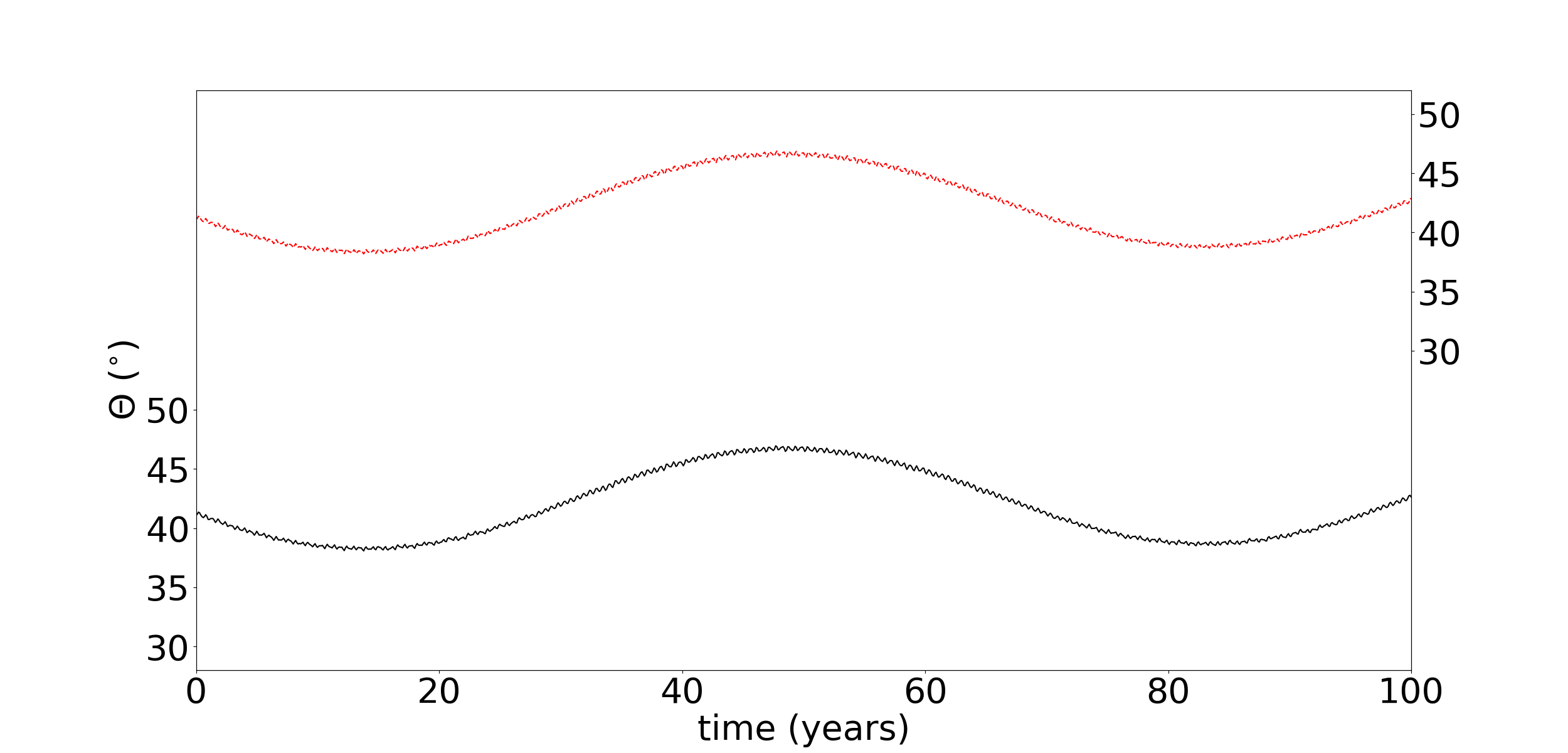}}
\caption{
Time evolution of  the (a) eccentricity and (c) azimuthal angle of a particle at $\Theta=41^{\circ}$ with the effects of Galatea with inclination $I = 0.0231^{\circ}$ (case~2,  in the bottom) and  $I = 0.0^{\circ}$ (case~1,  in the top). (b)  The  inclination  of the particle varies only for case~2. 
 }
\label{delta}
\end{figure}

 We  analyse  two cases, cases 1 and 2, as mentioned above. Figure~\ref{delta} shows the variation of the eccentricity, inclination and the azimuthal angle  for  a representative particle located in the Fraternit\'e arc in two different dynamical systems. In case~1  Galatea is in  equatorial and circular orbit and   in case~2 Galatea is in its nominal orbit ($e=0.00022$ and $I=0.0231^\circ $). In  both cases, 1 and 2, the  eccentricity of the particle varies between  $2.6 \times 10^{-4}$ to $4 \times 10^{-4}$. 
 The proximity to the 42:43 corotation inclined resonance due to Galatea affects the inclination of the particle,  reaching values up to $2.6 \times 10^{-4}$deg.

\begin{figure}
 \includegraphics[height=0.37\paperheight]{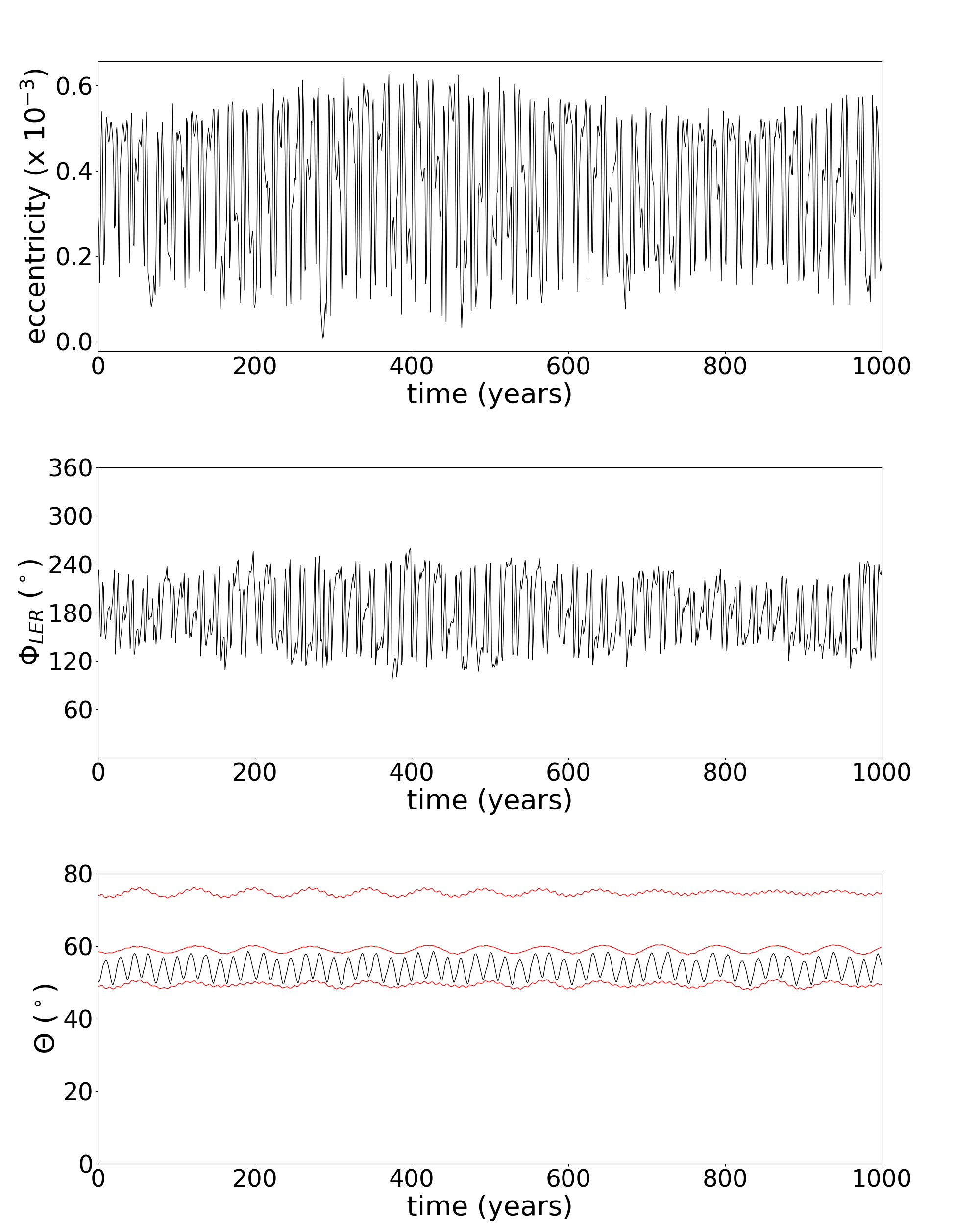}
 \caption{Time variation  of the eccentricity, the LER angle  and the azimuthal angle for a confined particle.}
 \label{conf}
 \end{figure}

We analyse the outcome of the numerical simulations by separating the particles into four different  groups. The  first group is composed  of  confined particles,  those particles that stay in the arc  under the effects of the Lindblad resonance with Galatea for the  whole  time of the numerical simulation. Figure~\ref{conf}  shows the variation of the eccentricity, the $\Phi_{LER}$ angle and the azimuthal  angle ($\Theta$) as a function of time for a particle confined in the arc.

\begin{figure}
 \includegraphics[height=0.37\paperheight]{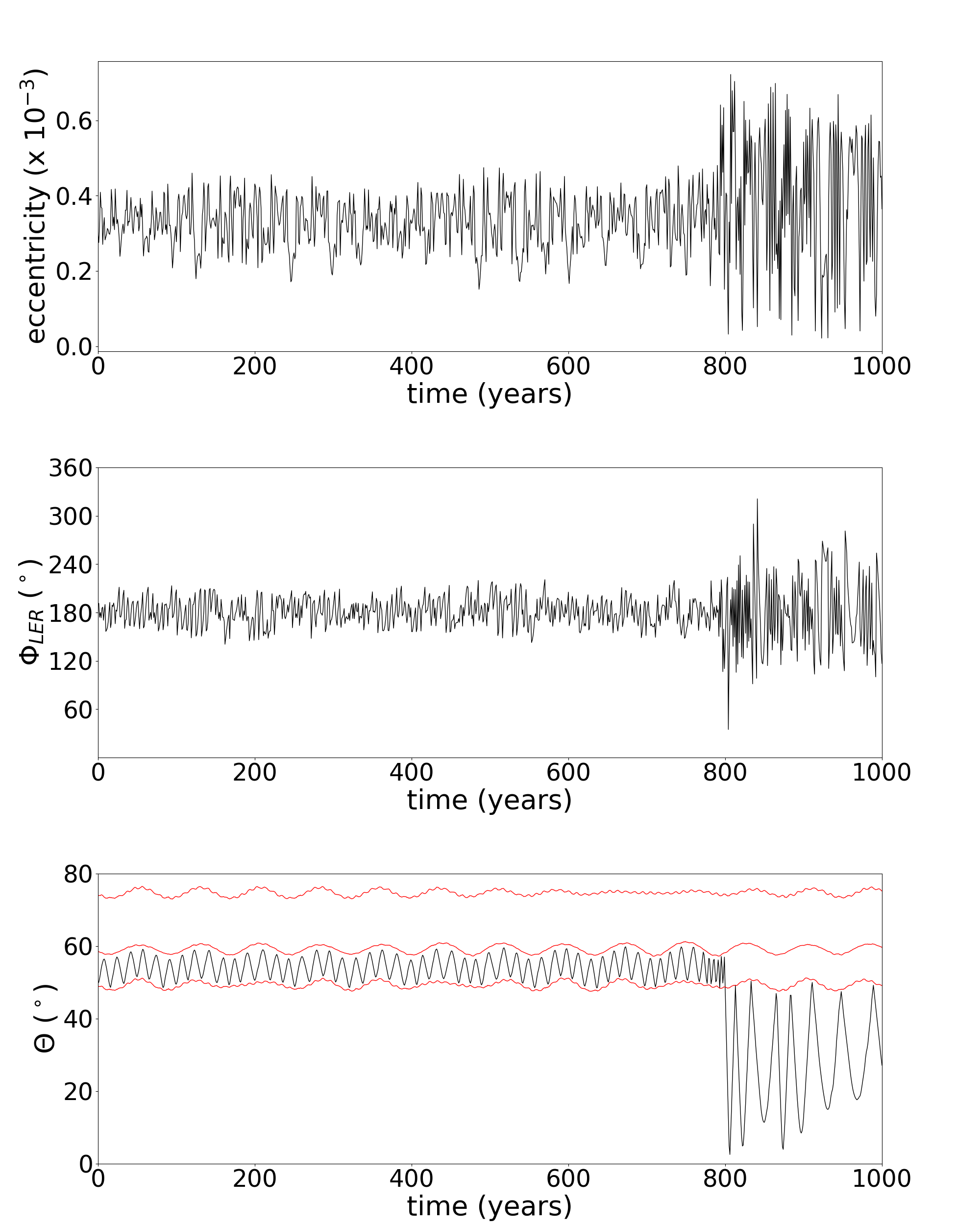}
 \caption{Time variation  of the eccentricity, the LER angle  and the azimuthal angle for a transient particle.}
 \label{trans}
 \end{figure}

\begin{figure}
 \includegraphics[height=0.37\paperheight]{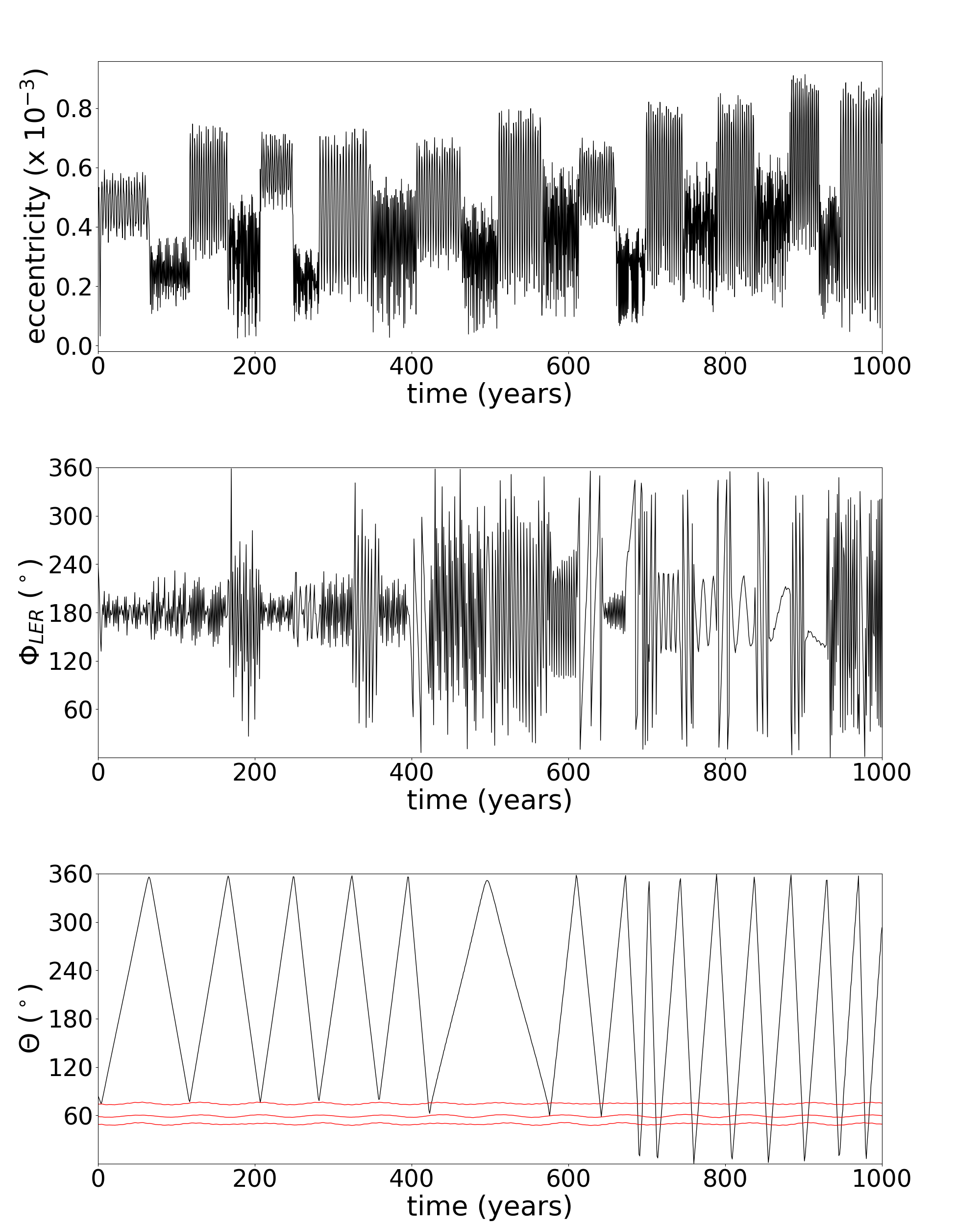}
 \caption{Time variation  of the eccentricity, the LER angle  and the azimuthal angle for a particle that leaves the arc.}
 \label{ring}
 \end{figure}

The second group is formed by transient particles, those particles   that travel between different arcs. These particles  can stay in the Lindblad resonance with Galatea, but  they change from one arc to another.  Figure~\ref{trans} shows the temporal variation of the  eccentricity,  $\Phi_{LER}$ and $\Theta$ angles  for a transient particle. 
The third group represents those particles that leave the arc and go to the Adams ring.  Figure~\ref{ring}  shows the eccentricity, $\Phi_{LER}$ and $\Theta$ angles as a function of time for a particle that leaves  the arc. The particle remains,  some time, in  the  Lindblad resonance. 
The last group is composed  of  those particles that collide with the co-orbital satellites.

\begin{figure}
 \includegraphics[height=0.6\paperheight]{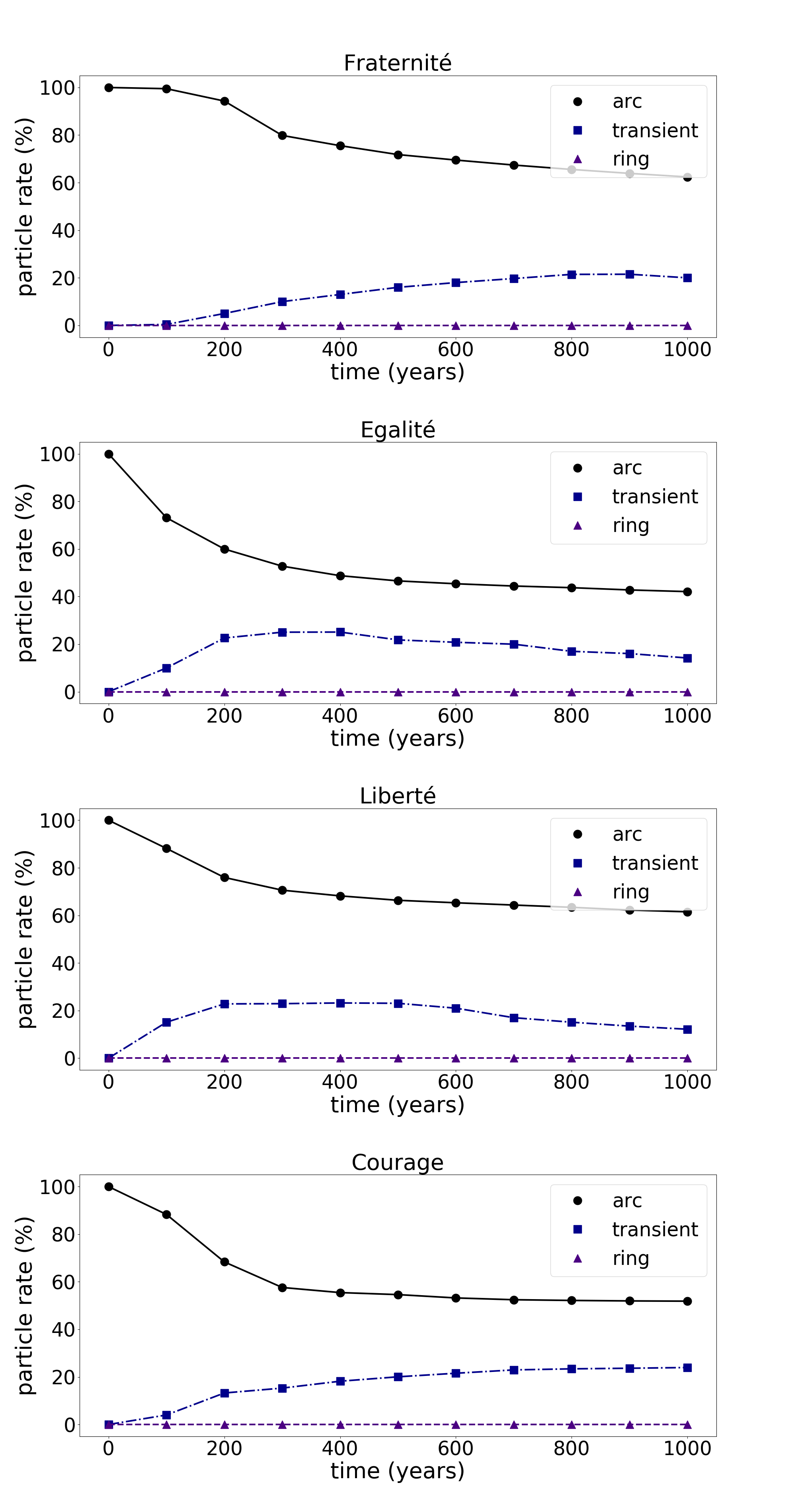}
 \caption{Percentage of confined, transient and those particles that go to the Adams ring as a function of time. The particles  are  under the effects of Galatea in an eccentric and  inclined  orbit. }
 \label{casecge}
 \end{figure}

 Figure~\ref{casecge} shows the percentage of confined and transient particles and those that leave the arc for case~2.  Both cases,  1 and 2,  have similar behaviour  concerning  the orbital evolution of the particles.  Although the evolution  of each arc (Figure~\ref{casecge}) presents  small  differences,  most of the particles  stay  in the four arcs for the whole time of integration, 1000~years. Some particles became transients, travelling between the arcs, but almost none of them goes to the Adams ring.

 The Egalit\'e and Libert\'e  arcs  are confined close to two co-orbital satellites, but it does not mean   that they are more stable than the Fraternit\'e and Courage arcs. The induced  eccentricity caused by Galatea  leads  to close encounters and collisions between the arc particles and the  co-orbital satellites.  

\section{Orbital Evolution under a Dissipative Force} \label{dissipative}

 In this system where the arc particles are confined in  the Lindblad resonance due to the satellite Galatea and in  the  corotation resonance due to  hypothetical co-orbital satellites we take a step further by including the effects of the solar radiation force \citep{Ma18}. 

\begin{figure}
 \includegraphics[height=0.15\paperheight]{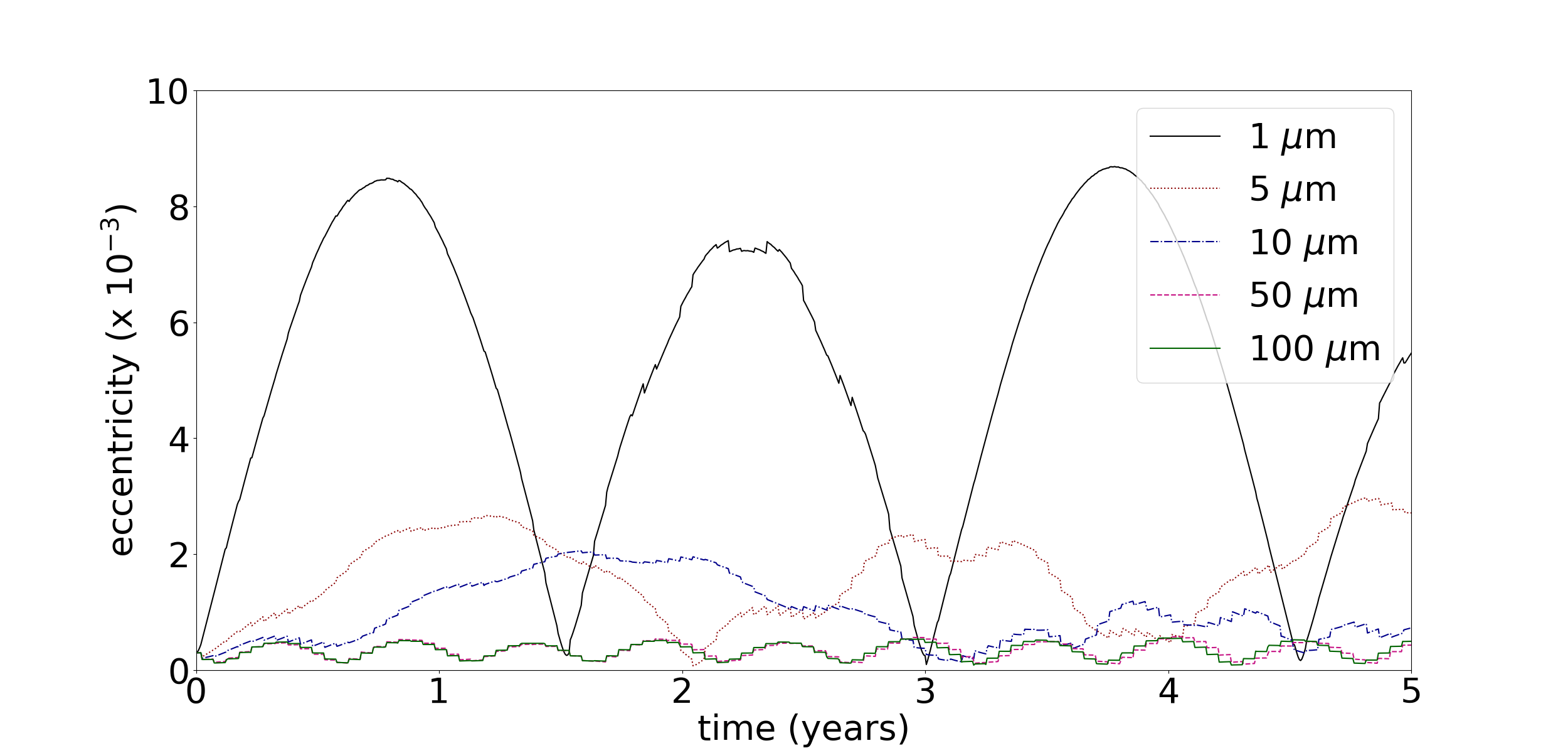} 
\caption{ Temporal variation of the eccentricity of  representative particles of different sizes located in the Fraternit\' e arc due to the radiation pressure.}
\label{exc}
 \end{figure}

The solar radiation force is composed  of  two components: the radiation pressure and the Poynting-Robertson components. 
The Poynting-Robertson component provokes a decrease in the semi-major axis of the particles in a timescale of thousands of years, while the radiation pressure component causes a variation in the eccentricity of the particles in a timescale of few years as shown in Figure~\ref{exc} for  a set  of  representative particles of different sizes ($1\mu$m, $5\mu$m, $10\mu$m, $50\mu$m and $100\mu$m in radius) at the Fraternit\' e arc. 

\begin{figure}
\subfigure[]{\quad \includegraphics[height=0.14\paperheight]{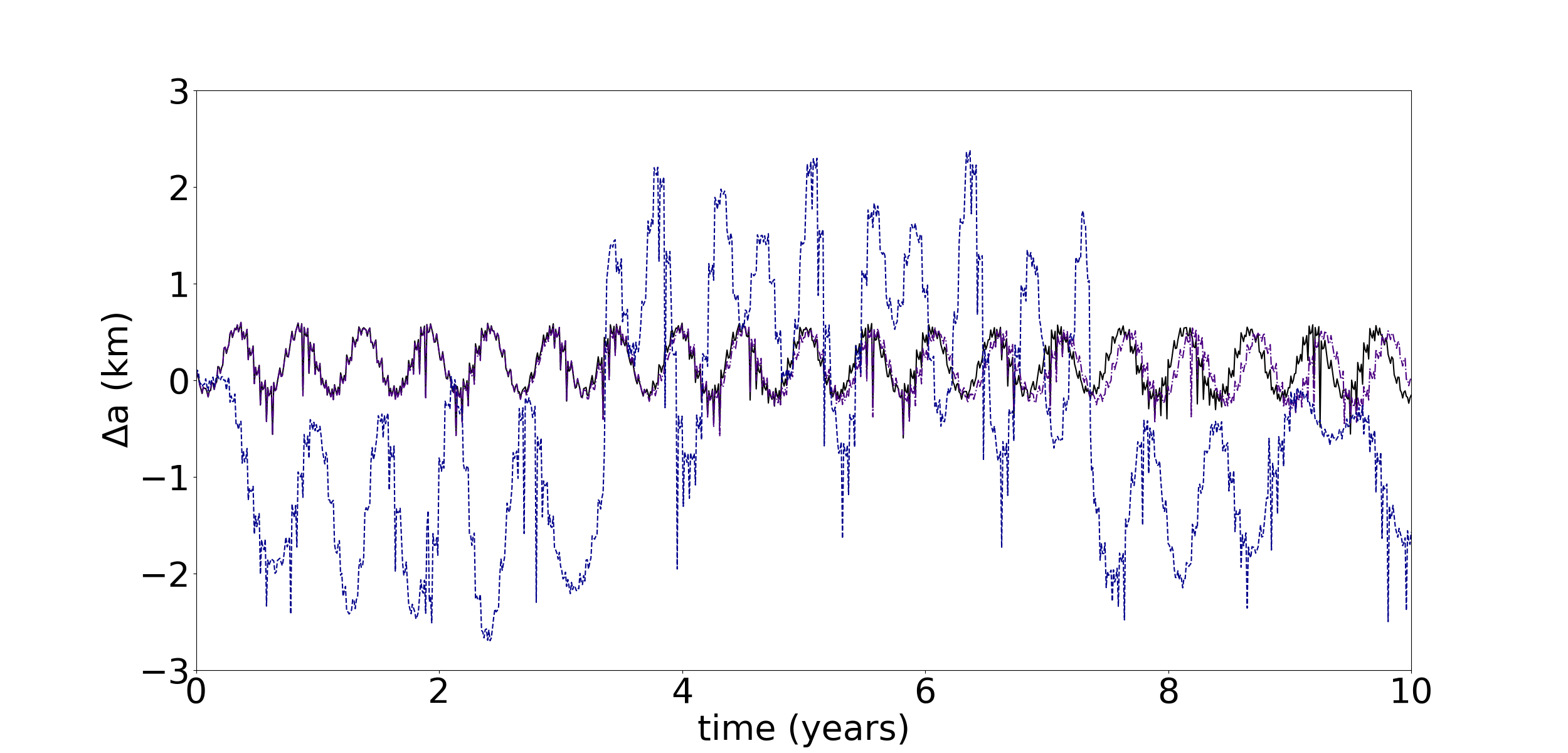}}
\subfigure[]{\quad \includegraphics[height=0.14\paperheight]{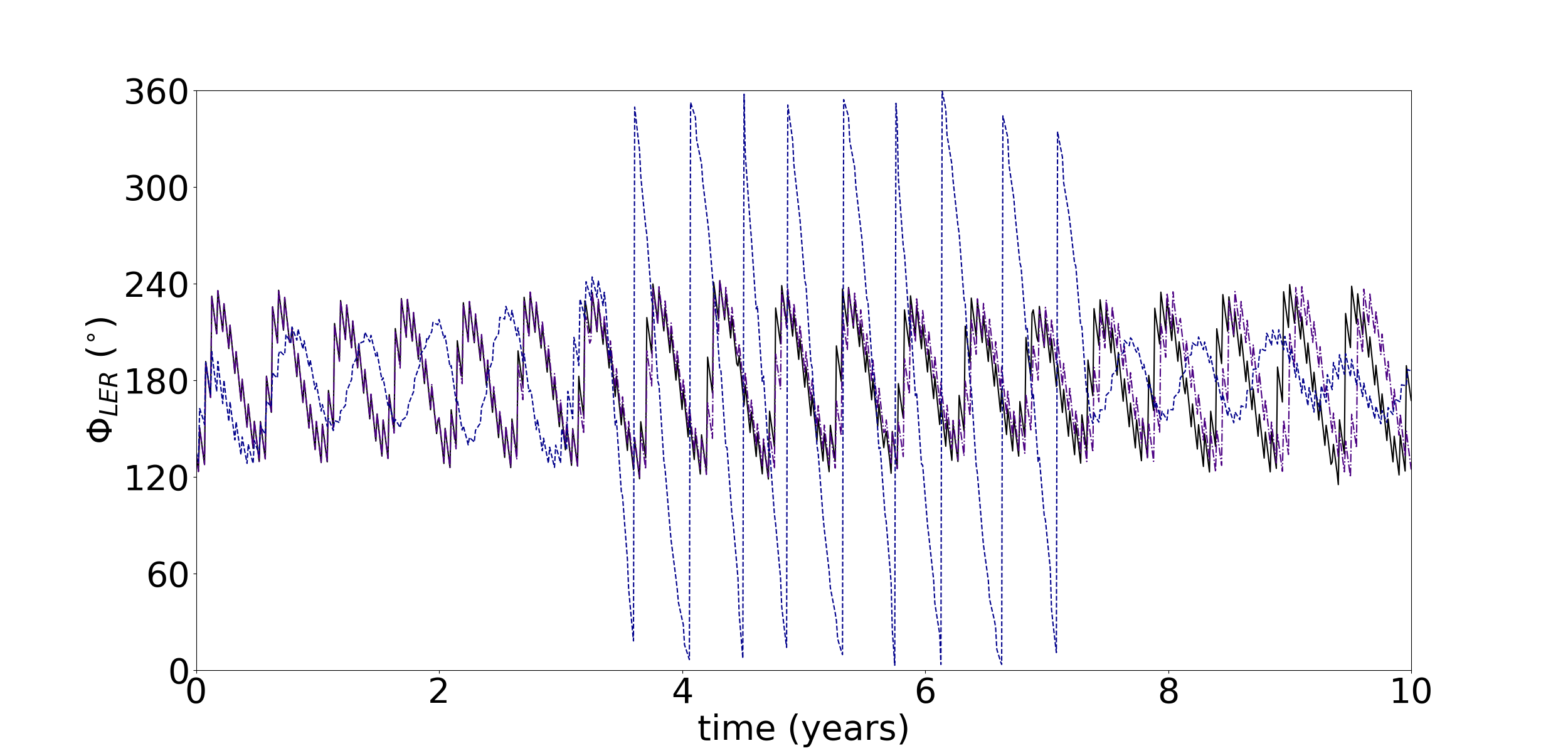}}
\subfigure[]{\quad \includegraphics[height=0.14\paperheight]{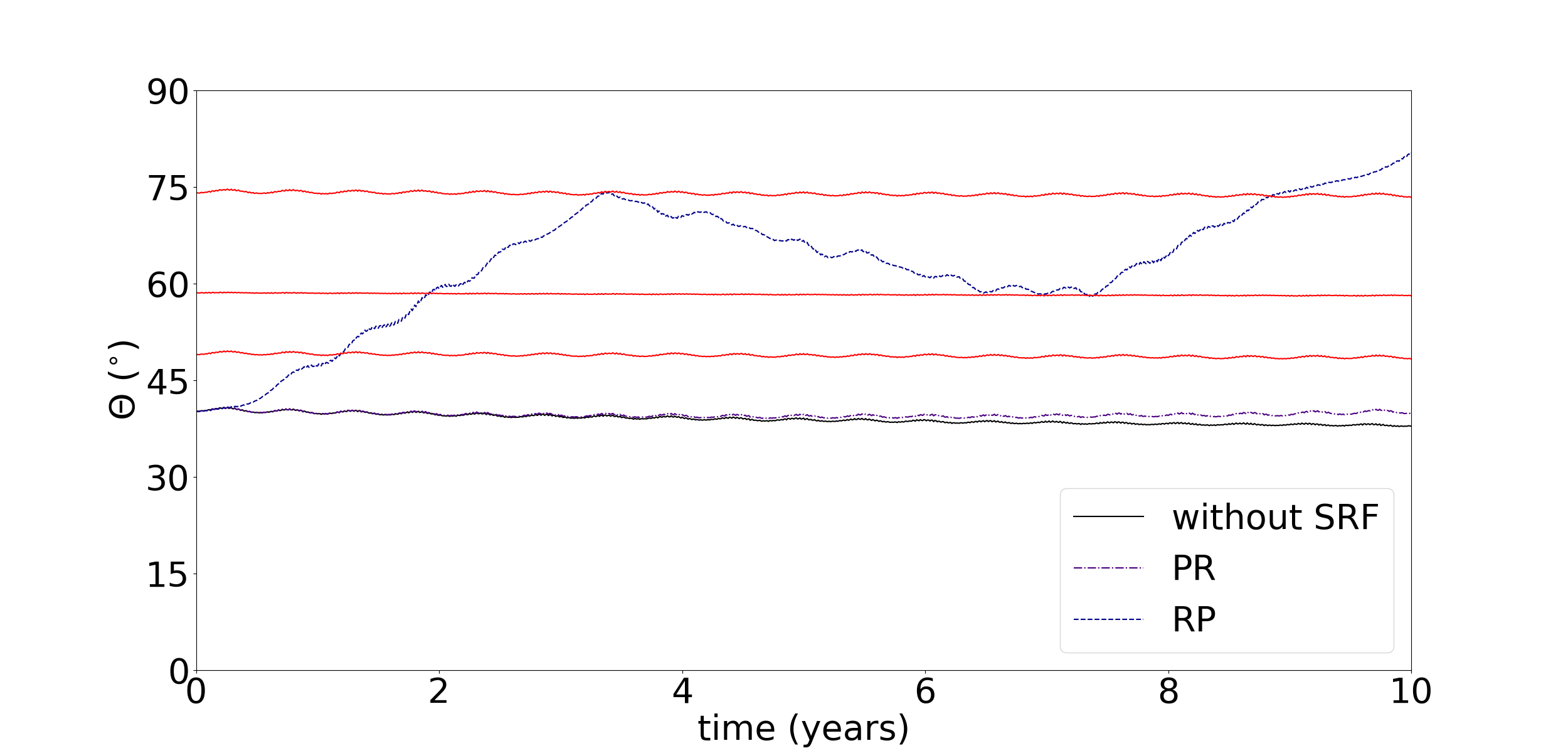}}
\caption{Time evolution of (a) the semi-major axis, (b) the LER angle and (c) the azimuthal angle ($\Theta=\lambda-\lambda_{S_1}$) of a particle initially at $\Theta=40^{\circ}$. The 10$\mu$m sized particle is under the effect of the gravitational force only,   the Poynting-Robertson component (PR) and  the radiation pressure component (RP).}
\label{deltas}
\end{figure}

  Figure~\ref{deltas} shows the temporal evolution of the  semi-major axis ($\Delta a$), the resonant angle and the azimuthal angle of a $10\mu$m sized  particle in the Fraternit\' e arc under the effects of  the Poynting-Robertson component,  the radiation pressure  component  and without any dissipative force.   Small variations in the semi-major axis of the particle are due to the proximity to the CIR with Galatea.  The  Poynting-Robertson component provokes small variations in the $\Phi_{LER}$ and $\Theta$ angles. This component does not remove particles from the resonance in 1000~years, the period of the numerical simulation.  
Variations in the eccentricity change the LER angle due to its constrain with the longitude of  the pericentre of the particle. Therefore,  the radiation pressure component  can remove particles from this resonance, as can be seen in Figure~\ref{deltas}. Furthermore, this component also provokes short-period oscillations in the semi-major axis of the particles, as discussed in \cite{Ma18}, which also  affect the $\Theta$ angle and  remove  the particles from the azimuthal confinement.

\begin{figure}
 \includegraphics[height=0.6\paperheight]{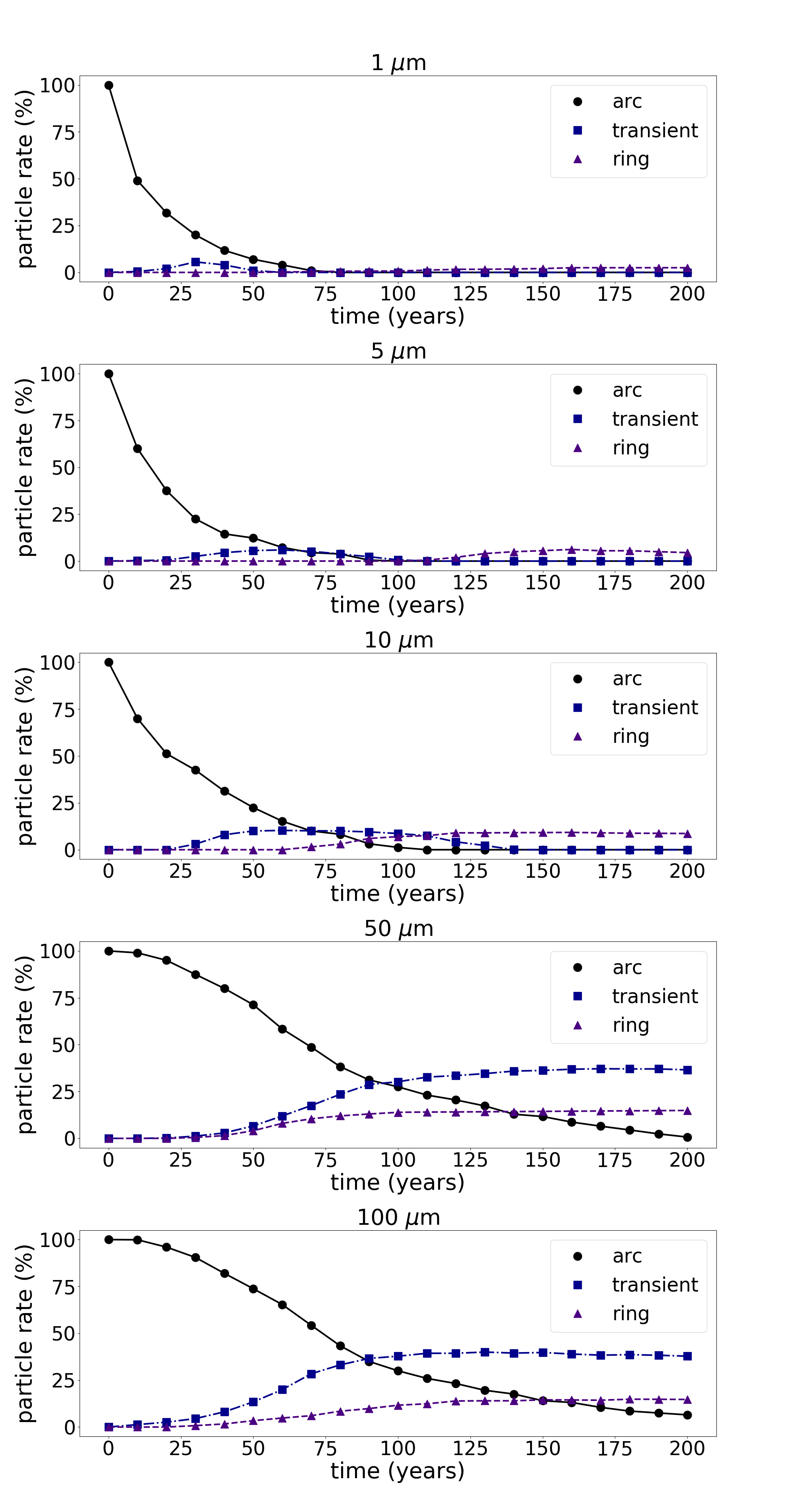}
 \caption{Percentage of confined, transient and  particles that go to the Adams ring as a function of time. The particles are initially  at the Fraternit\'e arc. }
 \label{fraternite}
 \end{figure}

We numerically simulated each arc composed of  a sample of 10000 particles,  2000 particles of sizes $1\mu$m, $5\mu$m, $10\mu$m, $50\mu$m and $100\mu$m in radius under the effects of the  solar radiation pressure.  
We analyse the percentage of confined and transient particles  and  those particles that leave the arcs and go to the Adams ring. Figure~\ref{fraternite} shows this percentage for particles of each size located in the Fraternit\'e arc. Particles at the Courag\'e, Libert\'e ,  and Egalit\'e arcs have similar behaviour.

By comparing  both systems, with (Fig.~\ref{fraternite})  and without (Fig.~\ref{casecge}) the solar radiation pressure,   we concluded that under the effects of this dissipative force  larger particles   live  up to  200~years.  Only under   the effects  of the  gravitational force more than $50\%$ of the particles remain  in the arc after 200~years.  As expected, the solar radiation force decreases the lifetime of the arcs,   smaller particles (1$\mu$m, 5$\mu$m and $10\mu$m in radius) collide with the  co-orbital satellites in a very short period of time. They do not became transient or leave  for the Adams ring.  The  population  of larger particles   contributes to the other arcs and also to the Adams ring: $50\mu$m and $100\mu$m sized particles can  become  transient or  move to the Adams ring.  The lifetime of the   smaller  arc particles  is  about  100~years.  

\begin{figure}
 \includegraphics[height=0.6\paperheight]{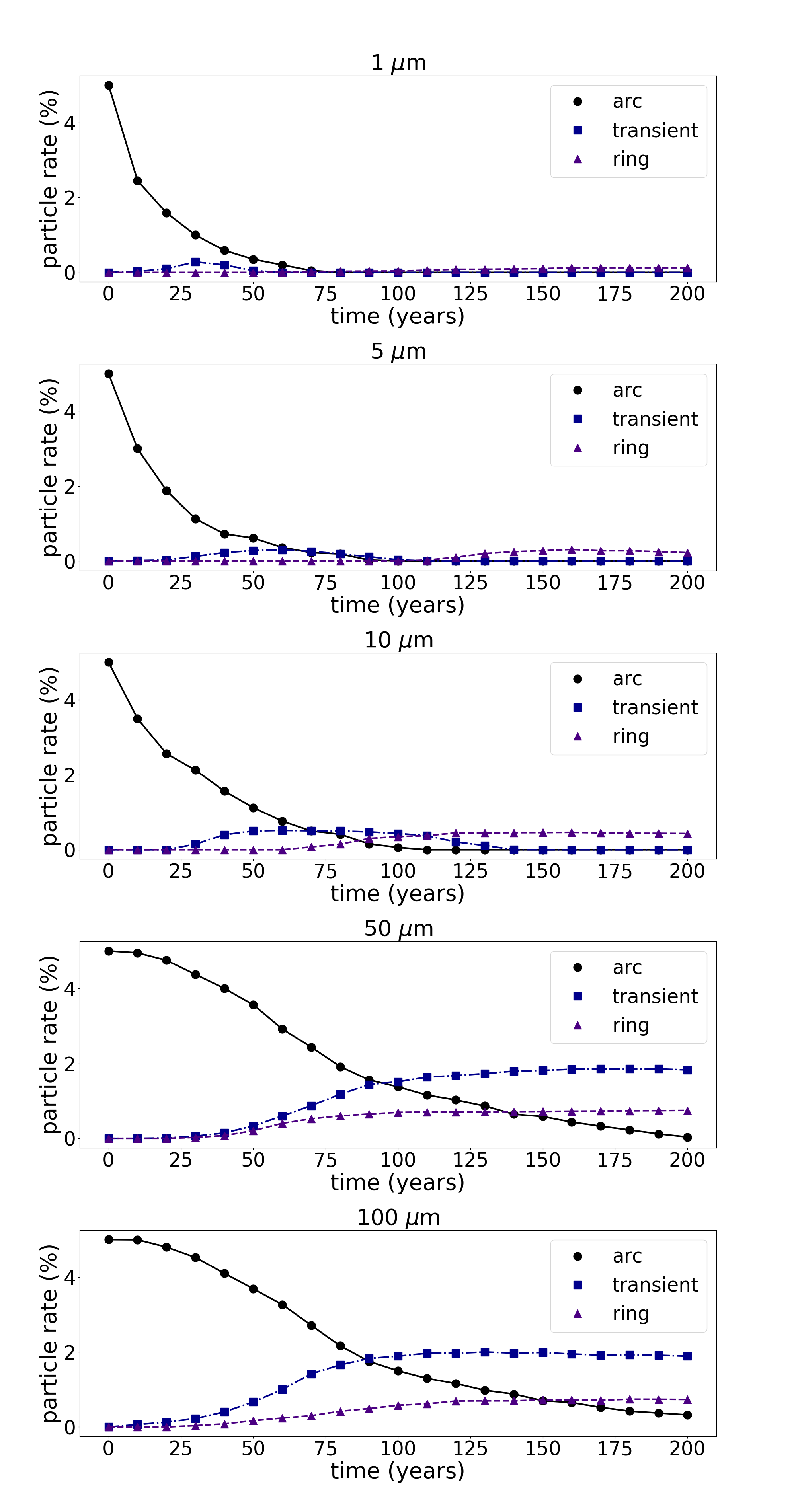}
 \caption{Percentage of confined, transient and  particles that go to the Adams ring as a function of time. The particles are initially  at the Fraternit\'e arc. Galatea is not present in the system. }
 \label{nosat}
 \end{figure}

  It is worth to point out that without the effects  of Galatea the system  has a similar  behaviour (Fig.~\ref{nosat} ).
 
   30~years ago  the arcs were imaged  by Voyager~2 spacecraft. Observations by the telescopes  showed that the arcs  changed in brightness \citep{Pa05,Sh13}.  Tables~\ref{T1} and  \ref{T2} show the percentage of  $\mu$m-sized particles  in each arc  and the  transient particles after 30 years, respectively.  
 $1\mu$m sized particles are quickly removed from the arcs, only $10\%$ of them can survive for 30~years.   The  Libert\'e arc can hold about $45\%$ of the $5\mu$m sized particles while the  Courage arc only $6\%$ of them.  More than $50\%$ of  the   larger particles ($50\mu$m and $100\mu$m in radius)  can survive in  all the arcs  after   this  period (30~years).
  
Table~\ref{T1}  shows that large particles  can stay in all  arcs after 30~years. These results can not explain the disappearance of the leading arcs,  Libert\'e and Courage,  unless the population of these arcs is formed by particles of different sizes.  Although  transient particles could  explain the  variation in the brightness of the arcs, the percentage of transient particles  is very small. 
 
From the  confinement model,  the  Fraternit\'e arc is closer to the larger co-orbital satellite ($S_1$), while the Egalit\'e arc is apart from $S_2$ for only $2^\circ$,  the  Libert\'e arc is apart from each co-orbital ($S_2$ and $S_3$) for about $7^\circ$ and  the Courag\'e arc is $12^\circ$  further from $S_4$.  The arcs are located in different configurations,  which may help to explain the small difference in the percentage of confined particles.

\begin{table}
\caption{ Percentage of $\mu $m-sized particles located in each arc after 30~years.}
\centering
\begin{tabular}{lcccc} \hline\hline
& Fraternit\'e  &  Egalit\'e &      Libert\'e & Courage  \\ 
\hline\hline
$1\mu$m   & $9$\% & $9$\%   & $12$\%  & $4$\% \\
$5\mu$m  & $20$\% & $25$\% &  $45$\%     & $6$\% \\
$10\mu$m  & $33$\% & $47$\% &  $52$\%     & $20$\% \\  
$50\mu$m  & $75$\% & $56$\% &  $63$\%     & $75$\% \\
$100\mu$m  & $82$\% & $65$\% &   $70$\%     & $80$\% \\                                             
\hline
\end{tabular}
\label{T1}
\end{table}


\begin{table}
\caption{ Percentage of transient particles  after  30~years. }
\centering
\begin{tabular}{lcccc} \hline\hline
              & Fraternit\'e  &  Egalit\'e &  Libert\'e & Courage  \\ 
\hline\hline
$1\mu$m   &   $7$\%  &  $3$\%   & $3$\%  & $2$\% \\
$5\mu$m  & $1$\%  & $5$\%   & $1$\%    & $2$\% \\
$10\mu$m  & $2$\% & $6$\%   &  --   & $1$\% \\                                             
$50\mu$m  & $2$\% & $2$\% &  $2$\%     & -- \\
$100\mu$m  & $6$\% & $4$\% &  $6$\%    & $1$\% \\                                               
 \hline
\end{tabular}
\label{T2}
\end{table}

\section{Dust Production } \label{production}

The arcs are temporary structures unless a source  provides the particles removed from them
through collisions or ejections. The co-orbital satellites could be the source (\cite{Pa05}
suggested that a moonlet immersed in the  Fraternit\'e arc can be its source) since they can be hit by
interplanetary dust particles (IDPs) that are diverted by the planet. These impacts usually  happen
at km/s, thus they are energetic enough to eject $\mu $m-sized particles from the surface of the target body.


The mass rate production ($M^{+}$) created by this process  depends on the flux of the impactors ($F_{\textrm imp}$),
the target  cross-section ($S$), and the ejecta yield ($Y$).
With these parameters the amount of dust produced is
$$M^{+} = F_{\textrm imp}~ S~ Y $$

\cite{Sf12} present a simple algorithm that allows the computation of the
mass rate production of  the  dust particles by this process. 
To compute $F_{\textrm imp}$ it was assumed an isotropic flux of IDPs at  Neptune's heliocentric distance of 
$1\times10^{-17}$ (km/m$^2$/s)  \citep{Po16}. However, the satellites are in fact hit by an enhanced flux due to gravitational focusing by the planet,  which
increases the IDPs velocity and their spatial density. In the region of the arcs, the effective flux is increased by a factor 30. 

The yield  ($Y$) is a measurement of the ejecta production efficiency, and it can be estimated using an 
 analytical expression presented in \citet{Ko01} which is a function of the impactor mass, its velocity and the proportion of silicate and ice 
on the surface of the satellite. For a typical impactor of $100\mu$m hitting a pure ice surface, we obtain $Y\sim 7-8\times10^{3}$.

\begin{figure}
\centering
\includegraphics[height=0.21\paperheight]{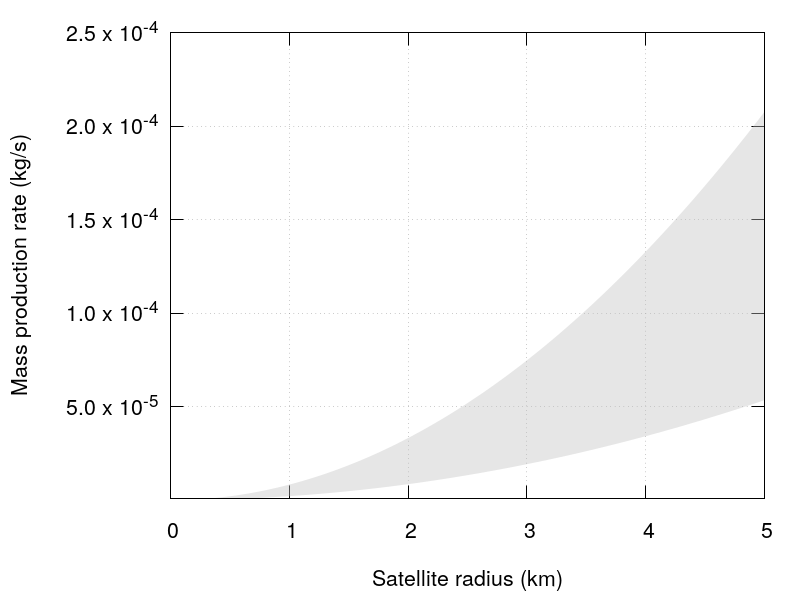}
\caption{The mass production rate by satellites with different radius.  
The grey area is due to  different values of the velocities  assumed  for the projectile: 2.3km/s and 3.0km/s.}
\label{impacts}
\end{figure}

With all these assumptions,  Figure~\ref{impacts} presents the mass production rate for different  sizes of satellites.  The grey area in Figure~\ref{impacts} comes from different values of the velocities assumed for
the projectile: 2.3km/s and 3.0km/s (\cite{Sf12} and \cite{Po16}, respectively). The IDPs population in the outer Solar System is poorly constrained. This range covers different models for the unfocused velocity of the projectiles: the value of 2.3km/s comes from the model proposed by \cite{Po16} where it is assumed that the origin of the flux at Neptune comes from the Edgeworth-Kuiper belt, while the higher value comes from a model adjusted with measurements made by New Horizons (\cite{Po16}). 

Due to the uncertainty in the parameters, mainly by the IDPs' flux which is poorly constrained at the heliocentric distance of 
Neptune,  the mass  generated by this process must be taken with one order of magnitude of uncertainty. Anyhow, 
even all particles being ejected with velocities large enough to overcome the escape velocity, the dust produced by the 
co-orbital satellites  does not contribute significantly to the arc population.

In order to  analyse  how efficient this dust production mechanism is, it was assumed that a small satellite, 5km in radius, 
produces $2 \times 10^{-4}$kg/s of dust (the maximum value derived from Fig.~\ref{impacts}).  It is also assumed that all  
particles generated on the surface of the  satellite survive for at least 1~year,  azimuthally   confined  in an arc of  
about $2^{\circ}$.  A crude estimative  of the optical depth of this   arc  can be obtained  from  the mass production ratio 
($2 \times 10^{-4}$kg/s).  Assuming that  the arc population  is  formed by particles with $1\mu$m in radius  we can calculate 
the total number of particles and the area (effective area) covered by them. The optical depth (the ratio between the effective 
area and the arc area)  is  $1 \times 10^{-5}$, smaller than the value $0.1$  presented in \cite{Po95}.  

Therefore, even considering the best 
scenario, the  larger satellite and the smallest arc, this mechanism of dust production can  not be  responsible 
for replenishing  the arcs.

\begin{figure}
 \includegraphics[height=0.25\paperheight]{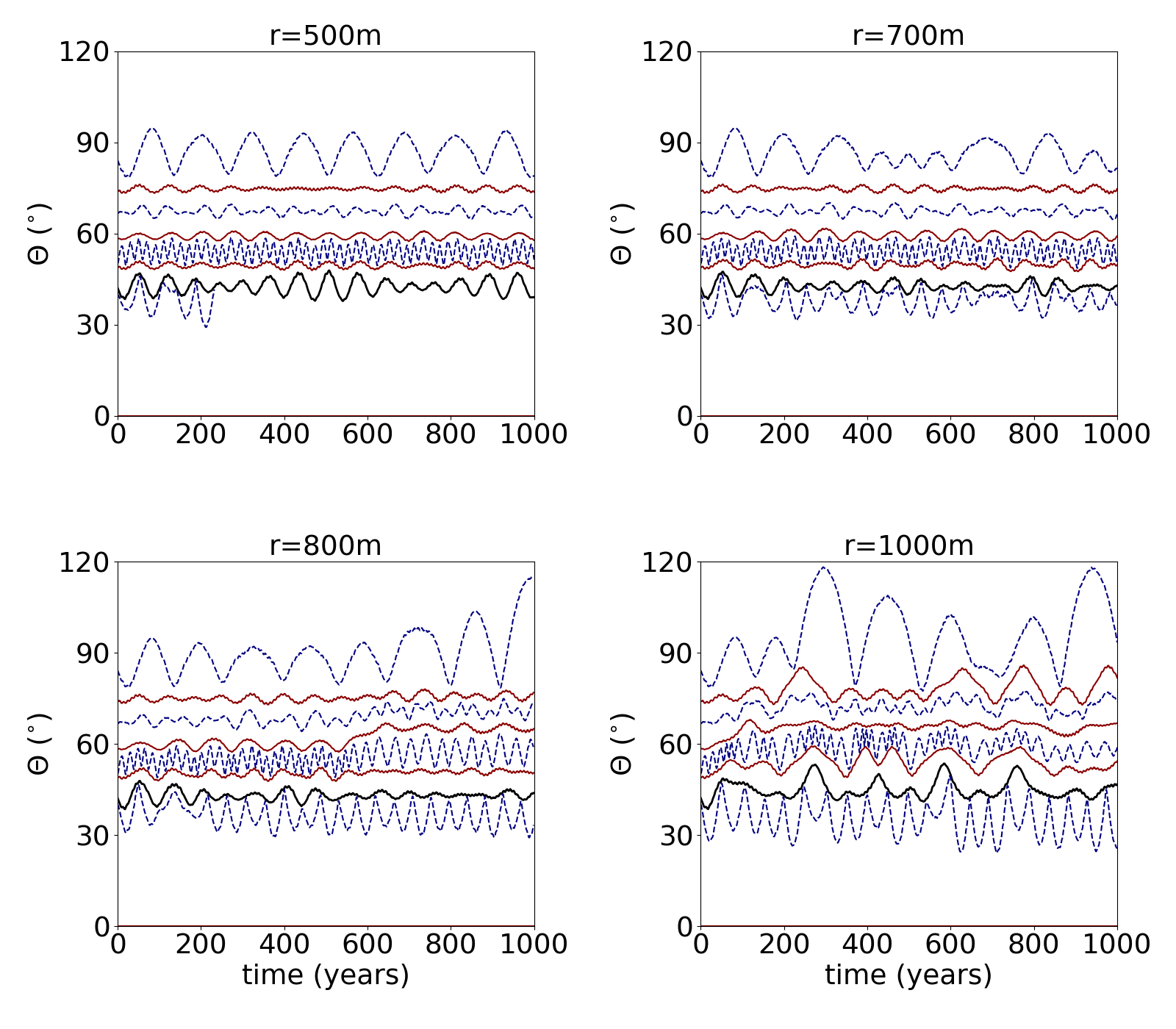}
 \caption{The co-orbitals satellites and the  arc particles disturbed by a moonlet located at the Fraternit\'e arc. Each figure shows  different  values of the radius ($r$) of the moonlet. }
 \label{moonletsp}
 \end{figure}

A sample of moonlets located in the arcs  can  disturb the orbital configuration of the co-orbital satellites, and consequently the arcs. In order to analyse the size of moonlets that can change the configuration of the system we modelled a system formed by four co-orbital satellites, a  moonlet located in the Fraternit\'e arc, and  a particle located in each arc. Figure~\ref{moonletsp}   shows the temporal variation of $\Theta$ as a function of time for the co-orbital satellites,  the moonlet and   the arc  particles.  We adopted moonlets with sizes 500m, 700m, 800m and 1000m in radius.  For a moonlet of radius  500m and 700m the configuration of the system is preserved, although the Fraternit\'e arc particle collided with the moonlet. For larger moonlets, the arc configuration changes significantly.  Therefore,  the  Fraternit\'e arc can have moonlets smaller than 700m in radius immersed on it.

A secondary population of dust production can be formed by dust produced by collisions between the arc particles onto the surface of the  co-orbital satellites or the immersed moonlets. The relative velocities of the dust  particles 
are about 60m/s for a collision between a  $1\mu$m sized particle and a co-orbital,  while a $100\mu$m sized particle hits the 
co-orbital satellites  at about 6.5m/s.  
These secondary collisions happen at velocities of a few m/s. Even dust particles could be released from the co-orbital satellites after these collisions, the amount is orders of magnitude smaller than direct impacts (collisions between IDPs and the  co-orbital satellites) that happen  at km/s. 
 
  Close to  the  Adams ring, an unnamed ring is co-orbital to the satellite Galatea \citep{Po95}, at about 61953km. Numerical simulations of the system formed by Neptune, the ring co-orbital particles and Galatea under the effects of the solar radiation force showed that although all small particles remained azimuthally confined (no collision with Galatea was detected), the smaller particles ($1\mu$m sized particles)  performed  radial oscillation of about 900km, while larger particles ($10\mu$m in radius)  oscillated about 120km.  The radial oscillations do not cross the inner edge of the Adams ring. Notwithstanding the unnamed ring can  influences the internal ring system of Neptune, it does not cause any change in the Adams ring and its arcs. Concerning the dust production, Galatea can  produce dust particles at a rate  of  approximately  a  hundred times faster than the hypothetical co-orbital satellites. However, most of the grains  leave the satellite at a speed that is below the escape velocity. Thus, it is unlikely that Galatea sustains the unnamed ring and helps to replenish the arcs with ejecta dust particles.

\section{Discussion}

 The   Adams ring arcs have been a challenger since their discovery in 1989. Confinement mechanisms have 
been proposed  to contain the spreading of the  arc particles.  Recent  data have shown that two arcs 
 changed  their intensity, while two others  are fading away. In this work we analyse the orbital evolution of the  arc
particles through the recent model which proposes that
co-orbital satellites (or large unseen  moonlets) can prevent the azimuthal spreading, while the satellite Galatea keeps their
radial spreading. 

The dynamical system is formed by an oblate Neptune, Galatea, dust  arc particles and four hypothetical small co-orbital satellites.  After the numerical simulations,   we divided the dust particles into four groups:  the  first group is formed by confined particles,  particles that stay in the same arc trapped in corotation and Lindblad resonances, the second group is  composed of transient particles (particles that change from one arc to another), the third group is formed by particles that leave the corotation resonance (the arcs) and  go  to the Adams ring, and the fourth group is composed of  those particles that collide with the co-orbital satellites. 

When only  the  gravitational force is acting in the system, about $50\%$ of the initial population of the arcs is present after 1000~years. However, the solar radiation force changes significantly this scenario.   Approximately  $25\%$ of the initial set of  smaller particles ($1\mu$m and  $5\mu$m  in radius) can  last  about 30~years. After 50~years all smaller particles collided with the  co-orbital satellites. Larger particles lived longer, but after 100~years only $25\%$ of the initial population is present in the arcs.   Apart from  the small differences, the four arcs have  similar dynamical behaviour.   None of them presents a  huge discrepancy that can, for example,  explain  the disappearance of one of them, unless they are formed by particles of different sizes.

The percentage of particles that stay in the arc is similar.
From our results the model proposed by \cite{Re14} does not explain why the two leading arcs are disappearing.  The   Courage arc would be  fading away  only  if it  is formed  by smaller particles.  The transient particles could help to explain the difference in the brightness of the arcs. However the
percentage of  them  is small.

Dust production due to inter-particle collisions onto the surface of large moonlets can be the source of dust particles, 
keeping the arcs in a steady state. By analysing the mass production rate in the best scenario, 
a large satellite (5km in radius)  closes to the  smallest arc, our results showed that 
the satellite  can not be the only source of the arcs.  Larger   moonlets  embedded in the arcs can  help to replenish the dust particles.  However,  we found  that  moonlets  larger than 0.7km in radius  change the configuration of  the co-orbital satellites  and the arcs, as proposed in the confinement model.  

Collisions may play an important role in the arcs. A recent paper by \cite{He19} studied the dynamics of multiple massive bodies in a corotation resonance. Their results showed that the bodies exchanged angular momentum and energy during the encounter which change  their orbits. They  argued that the exchange in the energy  may be similar to a collisional system, although a detailed  investigation is necessary.

Although no small satellite (less than 10km in radius) was detected in the Adams ring, it is
plausible to suppose that small bodies can be part of the ring population. Placed in  the right positions,
these satellites can trap  a number  of particles for a short period of time, until the solar radiation
force  removes  them from the arcs.  These co-orbital satellites may be the reminiscent of a large parent body that  brocke up,  and these  arcs are temporary features. This work is under investigation.

\section*{Acknowledgements}

The authors are grateful to the anonymous referee for helping to clarify many questions. The authors thank Fapesp (Procs~2016/24488-0, 2018/23568-6 and 2016/24561-0) and CNPq (Procs~309714/2016-8 and 305737/2015-5) 
for the financial support.  This study was financed in part by the Coordena\c c\~ao de Aperfei\c coamento de Pessoal de N\'\i vel Superior - Brasil (CAPES) - Finance Code 001.





\begin{thebibliography}{}
\bibitem[\protect\citeauthoryear{Burns et al.}{1979}] {Bu79} Burns, Joseph A and Lamy, Philippe L and Soter, Steven, 1979, Icarus, 40, 48
\bibitem[\protect\citeauthoryear{Chambers}{1999}] {Ch99} Chambers J.E., 1999, Monthly Notices of the Royal Astronomical Society, 304, 793
\bibitem[\protect\citeauthoryear{Dumas et al.}{1999}] {Du99} Dumas, Christophe and Terrile, Richard J and Smith, Bradford A and Schneider, Glenn and Becklin, EE, 1999, Nature, 400, 733
\bibitem[\protect\citeauthoryear{Foryta \& Sicardy}{1996}] {Fo96} Foryta, Dietmar W and Sicardy, Bruno, 1996, Icarus, 123, 123
\bibitem[\protect\citeauthoryear{Goldreich et al.}{1986}]{Go86} Goldreich, Peter and Tremaine, Scott and Borderies, Nicole, 1986, The Astronomical Journal, 92, 490
\bibitem[\protect\citeauthoryear{Hearn et al.}{2019}]{He19} A'Hearn, Joseph A and Hedman, Matthew M and El Moutamid, Maryame, 2019, The Astrophysical Journal, 882, 66
\bibitem[\protect\citeauthoryear{Hubbard et al.}{1986}]{Hu86} Hubbard, William B and Brahic, Andre and Sicardy, B and Elicer, LR and Roques, F and Vilas, F, 1986, Nature, 319, 636
\bibitem[\protect\citeauthoryear{Koschny \& Gr\"un}{2011}]{Ko01} Koschny, Detlef and Gr{\"u}n, Eberhard, 2011, Icarus, 154, 402
\bibitem[\protect\citeauthoryear{Lissauer}{1985}]{Li85} Lissauer, Jack J, 1985, Nature, 318, 544
\bibitem[\protect\citeauthoryear{Madeira et al.}{2018}]{Ma18} Madeira, G and Sfair, R and Mour\~ao, D C and Giuliatti Winter, S M, 2018, Monthly Notices of the Royal Astronomical Society, 475, 5474
\bibitem[\protect\citeauthoryear{Murray \& Dermott}{1999}]{MD99} Murray, Carl D and Dermott, Stanley F, 1999. Cambridge university press
\bibitem[\protect\citeauthoryear{Namouni \& Porco}{2002}]{Na02} Namouni, Fathi and Porco, Carolyn, 2002, Nature, 417, 45
\bibitem[\protect\citeauthoryear{Owen et al.}{1991}]{Ow91} Owen, WM and Vaughan, RM and Synnott, SP, 1991, The Astronomical Journal, 101, 1511
\bibitem[\protect\citeauthoryear{de Pater}{2005}]{Pa05} de Pater, Imke and Gibbard, Seran G and Chiang, Eugene and Hammel, Heidi B and Macintosh, Bruce and Marchis, Franck and Martin, Shuleen C and Roe, Henry G and Showalter, Mark, 2005, Icarus, 174, 263
\bibitem[\protect\citeauthoryear{Poppe}{2016}]{Po16} Poppe A. R., 2016, Icarus, 264, 369
\bibitem[\protect\citeauthoryear{Porco}{1991}]{Po91} Porco, Carolyn C, 1991, Science, 253, 995
\bibitem[\protect\citeauthoryear{Porco et al.}{1995}]{Po95} Porco, CC and Nicholson, PD and Cuzzi, JN and Lissauer, JJ and Esposito, LW, 1995, Neptune and Triton
\bibitem[\protect\citeauthoryear{Renner \& Sicardy}{2006}]{Re06} Renner, S and Sicardy, B, 2006,
Celestial Mechanics and Dynamical Astronomy, 94, 237
\bibitem[\protect\citeauthoryear{Renner et al.}{2014}]{Re14} Renner, S and Sicardy, B and Souami, D and Carry, B and Dumas, C, 2014, Astronomy \& Astrophysics, 563, A133
\bibitem[\protect\citeauthoryear{Sfair \& Giuliatti Winter}{2012}]{Sf12} Sfair, R and Giuliatti Winter, SM, 2012, Astronomy and Astrophysics, 543, 17
\bibitem[\protect\citeauthoryear{Showalter et al.}{2013}]{Sh13} Showalter, Mark R and de Pater, I and French, RS and Lissauer, JJ, 2013, AAS/Division for Planetary Sciences Meeting Abstract, 45
\bibitem[\protect\citeauthoryear{Showalter et al.}{2013}]{showalter2019seventh} Showalter, MR and de Pater, I and Lissauer, JJ and French, RS, 2019, Nature, 566, 7744
\bibitem[\protect\citeauthoryear{Sicardy et al.}{1991}]{Si91} Sicardy, Bruno and Roques, Fran{\c{c}}oise and Brahic, Andr{\'e}, 1991, Icarus, 89, 220
\bibitem[\protect\citeauthoryear{Sicardy et al.}{1999}]{Si99} Sicardy, B and Roddier, F and Roddier, C and Perozzi, E and Graves, JE and Guyon, O and Northcott, MJ, 1999, Nature, 400, 731
\end{thebibliography}


\bsp	
\label{lastpage}
\end{document}